\documentclass[%
reprint,
superscriptaddress,
 amsmath,amssymb,
 aps,
pre,
floatfix,
]{revtex4-2}

\usepackage{graphicx}%
\usepackage{amsmath}
\usepackage{bm}%
\usepackage[pdftex]{hyperref}%
\usepackage{xcolor}
\usepackage{ifthen}

\newcommand{\Eqref}[1]{\mbox{Eq.\hspace{0.25em}\eqref{#1}}}
\newcommand{\Eqsref}[1]{\mbox{Eqs.\hspace{0.25em}\eqref{#1}}}
\newcommand{\figref}[1]{\mbox{Fig.\hspace{0.25em}\ref{#1}}}

\newcommand{\secref}[1]{\mbox{\autoref{#1}}}

\newcommand{\iu}{\mathrm{i}\mkern1mu}

\newcommand{\vect}{\boldsymbol}
\newcommand{\diff}{\text{d}}

\newcommand{\mean}[1]{\langle #1 \rangle}

\newcommand{\muline}[1]{\underline{\smash{#1}\vphantom{T}}\vphantom{#1}}
\newcommand{\enthalpy}{h}%
\newcommand{\enthalpySpec}{\hat\omega}%

\newcommand{\Nc}{N}
\newcommand{\Vsys}{V_\mathrm{sys}}
\newcommand{\kBT}{k_\mathrm{B}T}
\newcommand{\cs}[1][]{\ifthenelse{\equal{#1}{}}{{\muline{c}}}{\muline{c}^{(#1)}}}
\newcommand{\cTot}{c_\mathrm{tot}}

\newcommand{\mus}{\muline{\mu}}

\newcommand{\phiIn}{\phi_\mathrm{in}}
\newcommand{\phiOut}{\phi_\mathrm{out}}
\newcommand{\pS}[1][]{\ifthenelse{\equal{#1}{}}{{c_\Nc}}{c_\Nc^{(#1)}}}
\newcommand{\p}[1][]{\ifthenelse{\equal{#1}{}}{{c}}{c^{(#1)}}}
\newcommand{\m}[1][]{\ifthenelse{\equal{#1}{}}{{\mu}}{\mu^{(#1)}}}

\newcommand{\mob}{\Lambda}
\newcommand{\mobR}{\mob_\mathrm{r}}
\newcommand{\mobP}{\mob_\mathrm{p}}
\newcommand{\mobA}{\mob_\mathrm{a}}
\newcommand{\sPa}{s^{(\mathrm{p})}}
\newcommand{\sAc}{s^{(\mathrm{a})}}

\newcommand{\sigmaA}{\sigma^{(\alpha)}}
\newcommand{\sA}{s^{(\alpha)}}
\newcommand{\sF}{s_\rightarrow}
\newcommand{\sB}{s_\leftarrow}

\newcommand{\sFA}{\sF^{(\alpha)}}
\newcommand{\sBA}{\sB^{(\alpha)}}

\newcommand{\kFA}{k_\rightarrow^{(\alpha)}}
\newcommand{\kBA}{k_\leftarrow^{(\alpha)}}
\newcommand{\KA}{K^{(\alpha)}}

\begin{document}

\title{The intertwined physics of active chemical reactions and phase separation}

\author{David Zwicker}
\email{david.zwicker@ds.mpg.de}
\affiliation{Max Planck Institute for Dynamics and Self-Organisation, Göttingen, Germany}%

\begin{abstract}
Phase separation is the thermodynamic process that explains how droplets form in multicomponent fluids. These droplets can provide controlled compartments to localize chemical reactions, and reactions can also affect the droplets' dynamics. This review focuses on the tight interplay between phase separation and chemical reactions, which originates from thermodynamic constraints. In particular, simple mass action kinetics cannot describe chemical reactions since phase separation requires non-ideal fluids. Instead, thermodynamics implies that passive chemical reactions reduce the complexity of phase diagrams and provide only limited control over the system's behavior. However, driven chemical reactions, which use external energy input to create spatial fluxes, can circumvent thermodynamic constraints. Such active systems can suppress typical droplet coarsening, control droplet size, and localize droplets. This review provides an extensible framework for describing active chemical reactions in phase separating systems, which forms a basis for improving control in technical applications and understanding self-organized structures in biological cells.
\end{abstract}

\maketitle

\tableofcontents


\section{Introduction}

Phase separation plays a crucial role for providing robust microstructures in
technology~\cite{Balachandar2010,Xu2021,Wang2019,Adler1988},
biology~\cite{Dignon2020,Alberti2019a,Brangwynne2009},
and even cooking~\cite{Mathijssen2022}.
In many cases, these structures provide a basis for controlling chemical reactions or are themselves affected by reactions.
Examples include chemically fueled assemblies~\cite{Spath2021, Bazant2017, Balazs2007, Tran-Cong1996} and membrane-less compartments that control reactions in biological cells~\cite{Oflynn2021,Nakashima2019, Sokolova2013,Saurabh2022}, which are also often regulated using chemical modifications of the involved biomolecules~\cite{Saurabh2022,Nakashima2018, Soeding2019, Snead2019, Hondele2020, Kirschbaum2021}.
In fact, chemically controlled droplets might explain how cells originated at the origin of life~\cite{Ghosh2021a,Nakashima2021,Donau2020}.
All these examples have in common that phase separation and chemical reactions are strongly intertwined.

To illustrate the connection between phase separation and chemical reactions, let us consider a fluid comprising two molecular species, $A$ and~$B$, which can phase separate and interconvert, $A \rightleftharpoons B$.
In the simplest case, phase separation leads to two homogeneous phases whose compositions are respectively characterized by the concentration pairs $\{c_A^{(1)}, c_B^{(1)}\}$ and $\{c_A^{(2)}, c_B^{(2)}\}$. %
The concentration ratios between the two compartments define partition coefficients %
\begin{align}
	\label{eqn:partition_coefficient_simple}
	P_A &= \frac{c_A^{(1)}}{c_A^{(2)}}
& \text{and} &&
	P_B &= \frac{c_B^{(1)}}{c_B^{(2)}}
	\;,
\end{align}
with $P_A \neq 1$ or $P_B \neq 1$ in a phase separated state.
Conversely, the chemical equilibrium of the conversion reaction in each phase is described by equilibrium constants, which are ratios of the concentrations of products and substrates,
\begin{align}
	\label{eqn:reaction_equilibrium_ratio_simple}
	K^{(1)} &= \frac{c_B^{(1)}}{c_A^{(1)}}
& \text{and} &&
	K^{(2)} &= \frac{c_B^{(2)}}{c_A^{(2)}}
	\;.
\end{align}
Combining \Eqsref{eqn:partition_coefficient_simple} and \ref{eqn:reaction_equilibrium_ratio_simple} implies
\begin{align}
	\label{eqn:relationship_simple}
	\frac{P_A}{P_B} = \frac{K^{(2)}}{K^{(1)}}
	\;,
\end{align}
which demonstrates that phase separation and chemical reactions are intimately linked.

The main purpose of this review is to reveal consequences of the connection between phase separation and chemical reactions, and show how non-equilibrium conditions can be used to circumvent them.
We start by introducing a general theory for multi-component fluids in \secref{sec:theory}, which culminates in a full dynamical description and a generalization of \Eqref{eqn:relationship_simple}.
We then summarize known consequences of reactions in phase separating systems for binary fluids (\secref{sec:binary}) and ternary systems (\secref{sec:ternary}).
However, there are many unexplored aspects and we list some future challenges in \secref{sec:future}.

\section{Theory of multicomponent fluids}
\label{sec:theory}

We consider a regular solution consisting of $N$ different component $X_1$, $X_2$, \ldots, $X_N$.
The state of a homogeneous system is then fully specified by its volume $V$, temperature~$T$, and the particle counts~$N_i$ for all species $i=1,\ldots,N$.
If the interactions at the system's boundary are negligible, the free energy~$F$ of the system is given by $F(\{N_i\},V,T)=V f(\cs,T)$, where the concentrations $c_i = N_i/V$ define the composition $\cs=(c_1, \ldots, c_N)$.
We split the free energy density~$f$ into the ideal entropy of mixing and the enthalpic contributions~$\enthalpy$,
\begin{equation}
	\label{eqn:free_energy_many}
	f(\cs,T) = 
		\kBT \left[
			\sum_{i=1}^{N} c_i \ln\left(\frac{c_i}{\sum_j c_j}\right)
			+ \enthalpy(\cs,T)
		\right]
	\;,
\end{equation}
where $k_\mathrm{B}$ is Boltzmann's constant.
The enthalpy density~$\enthalpy$ captures internal degrees of freedom of the molecules as well as short-ranged interactions, including various protein interactions~\cite{Dignon2020} and complex  coacervation~\cite{Overbeek1957}.
A special case is ideal solutions, where interactions are absent and all components have constant internal energy~$w_i$, so $\enthalpy(\cs,T)=\sum_i c_i w_i(T)$.
Another example is the Flory-Huggins free energy, $\enthalpy(\cs, T) = \sum_{i,j} c_ic_j\,\chi_{ij}(T)$~\cite{Flory1942,Huggins1941,Mao2018}, where the Flory-parameters~$\chi_{ij}$ encode pairwise interactions.
In all cases, free energies~$F$ imply chemical potentials $\mu_i = (\partial F/\partial N_i)_{V,T,N_{j\neq i}}$, which read
\begin{align}
	\label{eqn:chemical_potential_homogeneous}
	\mu_i &= \kBT \left[ 
		 \ln\left(\frac{c_i}{\sum_j c_j}\right) + \enthalpySpec_i(\cs,T) 
	   \right]
	  \;,
\end{align}
where $\enthalpySpec_i = \partial \enthalpy/\partial c_i$ are specific enthalpies.
Similarly, we obtain the pressure~$\Pi=-(\partial F/\partial V)_{T,N_i} = \sum_i c_i \mu_i - f$, 
\begin{align}
	\label{eon:pressure_many}
	\Pi &=  \kBT\Biggl[
		 \sum_{i=1}^N c_i \, \enthalpySpec_i(\cs,T) 
		- \enthalpy (\cs, T)
		\Biggr]
	\;.
\end{align}
Finally, we could determine entropies from derivatives with respect to temperature~$T$, but we will  focus on isothermal systems with constant $T$ for simplicity.
Consequently, only the intensive quantities $\mu_i$ and $\Pi$ govern chemical reactions and phase separation, which we will first discuss separately.

\subsection{Chemical reactions}
\label{sec:chemical_reactions}

We consider a system with $M$ chemical reactions enumerated by $\alpha=1,\ldots, M$.
Each reaction is of the form
\begin{align}
	\sum_{i=1}^N \sigmaA_{\rightarrow,i} \,X_i \rightleftharpoons \sum_{i=1}^N \sigmaA_{\leftarrow,i} \,X_i
	\;,
\end{align}
where $\sigmaA_{\rightarrow,i}$ and $\sigmaA_{\leftarrow,i}$ denote the stoichiometric coefficients of  reactants and  products, respectively.
They can be combined in the stoichiometry matrix
\begin{align}
	\sigmaA_i = \sigmaA_{\leftarrow,i} - \sigmaA_{\rightarrow,i}
	\;,
\end{align}
where $\alpha=1,\ldots,M$ and $i=1,\ldots, N$.
Chemical reactions conserve mass, $\sum_i\sigmaA_i m_i = 0$, where $m_i$ denote molecular masses.
They also imply conserved quantities $\psi_\beta = \sum_i q^{(\beta)}_i c_i$, where $q^{(\beta)}_i$  are linearly independent vectors in the cokernel of the stoichiometric matrix, $\sum_i q^{(\beta)}_i\sigmaA_i = 0$~\cite{Avanzini2021}.
For instance, the concentration $c_i$ of component $X_i$ is a conserved quantity if it does not participate in any reaction.
Another example is simple conversion reactions, $X_i \rightleftharpoons X_j$, where $c_i + c_j$  is conserved.

\subsubsection{Reaction equilibrium}
Thermodynamic equilibrium of a reaction is reached when the chemical potentials balance, 
\begin{align}
	\label{eqn:chemical_equilibrium}
	\sum_{i=1}^N \sigmaA_i \mu_i &= 0
	\;.
\end{align}
The associated  equilibrium ratio~$\KA= \prod_i c_i^{\sigmaA_i}$ can then be expressed using \Eqref{eqn:chemical_potential_homogeneous}.
A particularly simple form emerges when reactions conserve particle counts, $\sum_i\sigmaA_i = 0$, e.g., because all species have equal molecular mass~$m_i$.
In this case, 
\begin{align}
	\label{eqn:reaction_equilibrium_ratio}
	\KA(\cs) &
	= \exp\left[-\sum_{i=1}^N \sigmaA_i \enthalpySpec_i(\cs)\right]
	\;,
\end{align}
which may depend on the composition~$\cs$.
For an ideal solution, $\enthalpySpec_i(\cs) = w_i $, this reduces to the familiar equilibrium constant,
\begin{align}
	\KA_\mathrm{ideal} &
	=  \exp\left[-\sum_{i=1}^N\sigmaA_i w_i\right]
	\;,
\end{align}
which is constant because it only depends on the stoichiometries and the internal energies~$w_i$.

\subsubsection{Reaction kinetics}
The evolution of a composition~$\cs=(c_1, c_2, \ldots, c_N)$ toward equilibrium is described by $\partial_t c_i = s_i$, where
\begin{align}
	\label{eqn:reaction_flux_species}
	s_i
	  = \sum_{\alpha=1}^M \sigmaA_i \sA
\end{align}
is the total rate of production of component~$i$.
Here, the net flux~$\sA$ of reaction~$\alpha$ can be split into a forward direction, $\sFA$, and a backward direction, $\sBA$,
\begin{align}
	\sA = \sFA - \sBA
	\;,
\end{align}
which must balance in equilibrium, $\sA=0$.

\paragraph{Thermodynamic constraints of the rates}
Thermodynamics does not only impose a vanishing rate when the chemical potentials are balanced, but also the stronger condition
\begin{align}
	\label{eqn:detailed_balance_of_rates}
	\frac{\sFA}{\sBA} 
		&= \exp\left[-\frac{\sum_i \sigmaA_i \mu_i}{\kBT}\right]
	\;,
\end{align}
which is known as \emph{detailed balance of the rates}~\cite{Weber2019, Julicher1997}.
This condition ensures that the reaction proceeds in the forward direction, $\sFA > \sBA$, when the products are energetically favored, $\sum_i \sigmaA_{\leftarrow,i}\,\mu_i < \sum_i \sigmaA_{\rightarrow,i}\,\mu_i $, which is a consequence of the second law of thermodynamics.

\paragraph{Transition state theory}
The constrain~\eqref{eqn:detailed_balance_of_rates} determines the ratio of forward to backward flux, but it does not determine the magnitude of either.
Indeed, such rates cannot be determined from thermodynamics, so kinetic models are necessary.
One of the simplest model is \emph{transition state theory}, where an unstable transition state forms transiently during the reaction. 
The associated fluxes can be expressed as~\cite{Pagonabarraga1997,Hanggi1990}
\begin{subequations}
\label{eqn:reaction_fluxes_passive}
\begin{align}
	\sFA &= k_\alpha \exp\left(\frac{\sum_i \sigmaA_{\rightarrow,i} \, \mu_i}{\kBT}\right)
	\qquad\text{and}
\\
	\sBA &= k_\alpha \exp\left(\frac{\sum_i \sigmaA_{\leftarrow,i} \, \mu_i}{\kBT}\right)
	\;,
\end{align}
\end{subequations}
where the positive pre-factor~$k_\alpha$ might depend on composition.
The net flux of reaction~$\alpha$ thus reads
\begin{align}
	\label{eqn:reaction_flux_total}
	\sA &= k_\alpha \left[
		\exp\left(\frac{\sum_i \sigmaA_{\rightarrow,i} \, \mu_i}{\kBT}\right)
		- \exp\left(\frac{\sum_i \sigmaA_{\leftarrow,i} \, \mu_i}{\kBT}\right)
	\right]
	\;,
\end{align}
where the square bracket denotes the chemical reaction force~\cite{Bauermann2021}.
This flux can be expressed in the form of mass-action kinetics,
\begin{align}
	\sA &= \kFA \prod_{i=1}^N (c_i)^{\sigmaA_{\rightarrow,i}} - \kBA \prod_{i=1}^N (c_i)^{\sigmaA_{\leftarrow,i}}
	\;,
\end{align}
with the respective forward and backward rates
\begin{subequations}
\begin{align}
	\kFA &=k_\alpha \,
		\cTot^{-\sum_i\sigmaA_{\rightarrow,i}} \,
		\prod_{i=1}^N e^{w_i  \, \sigmaA_{\rightarrow,i} } 
\qquad \text{and} \\
	\kBA &=k_\alpha \,
		\cTot^{-\sum_i\sigmaA_{\leftarrow,i}} \,
		\prod_{i=1}^N e^{w_i  \, \sigmaA_{\leftarrow,i}}
	\;,
\end{align}
\end{subequations}
where $\cTot = \sum_i c_i$.
Simple mass-action kinetics with constant rates emerges when $k_\alpha$, $\cTot$, and $\enthalpySpec_i$ are constant, e.g., in an incompressible, ideal fluid. 
In contrast, chemical reactions in non-ideal solutions generally deviate from mass-action kinetics.
Close to equilibrium, \Eqref{eqn:reaction_flux_total} can be linearized in the chemical potentials to obtain
$\sA = -\mobR \sum_i \sigmaA_i \mu_i$, where $\mobR = k_\alpha / \kBT$ is the reaction mobility.
This form is thermodynamically consistent and captures the qualitative kinetics of chemical reactions in non-ideal solutions.

\subsubsection{Active systems}
So far we have considered closed systems, which relax to equilibrium and can thus be classified as \emph{passive}.
In contrast, \emph{active} systems are kept away from equilibrium, which is only possible if they are open, so particles can exchange with the environment.
In the simplest case, one or more species are coupled to a particle reservoir, or \emph{chemostat}, so their chemical potentials are kept constant at the system boundary.
There are then cases where the chemical equilibrium, \Eqref{eqn:chemical_equilibrium}, cannot be satisfied for all reactions and detailed balance is thus broken.
In such cases, chemical reaction networks can display complex dynamics, including oscillations~\cite{Nicolis1977}, which are sustained by the chemostatted species acting as a fuel.

We  consider open systems where~$\tilde N$ additional fuel molecules $\tilde X_j$ are coupled to chemostats, while the internal components~$X_i$ cannot cross the system's boundary.
The chemostatted molecules participate in driven reactions,
\begin{align}
	\sum_{i=1}^N \sigmaA_{\rightarrow,i} \, X_i  
		+ \sum_{j=1}^{\tilde N} \tilde \sigma^{(\alpha)}_{\rightarrow,j} \, \tilde X_j
	\rightleftharpoons 
	\sum_{i=1}^N \sigmaA_{\leftarrow,i} \, X_i
		+ \sum_{j=1}^{\tilde N} \tilde \sigma^{(\alpha)}_{\leftarrow,j} \, \tilde X_j
	\;,
\end{align}
where $\tilde\sigma^{(\alpha)}_{\rightarrow,j}$ and $\tilde\sigma^{(\alpha)}_{\leftarrow,j}$ are their stoichiometric coefficients. %
For simplicity, we assume that the overall density of fuel~$\tilde X_j$ is low and that they do not interact with the other molecules~$X_i$.
In this case, chemical potentials of $X_i$ are unchanged and the only effect of the fuel molecules is to supply chemical energy via their chemical potentials $\tilde\mu_j$, which are kept constant via the chemostat.
For each reaction~$\alpha$, we can thus define the chemical energies of the chemostatted reactants and products,
\begin{align}
	\label{eqn:mu_fuel_multiple}
	\tilde\mu^{(\alpha)}_\rightarrow &= \sum_{j=1}^{\tilde N} \tilde \sigma^{(\alpha)}_{\rightarrow,j} \,\tilde\mu_j
& \text{and}&&
	\tilde\mu^{(\alpha)}_\leftarrow &= \sum_{j=1}^{\tilde N} \tilde\sigma^{(\alpha)}_{\leftarrow,j}\,\tilde\mu_j
	\;.
\end{align}
Using transition state theory, the reaction fluxes are
\begin{subequations}
\label{eqn:reaction_fluxes_active}
\begin{align}
	\sFA &= k_\alpha \exp\left(\frac{\tilde\mu^{(\alpha)}_\rightarrow + \sum_i \sigmaA_{\rightarrow,i} \, \mu_i}{\kBT}\right) \qquad \text{and}
\\
	\sBA &= k_\alpha \exp\left(\frac{\tilde\mu^{(\alpha)}_\leftarrow + \sum_i \sigmaA_{\leftarrow,i} \, \mu_i}{\kBT}\right)
	\;.
\end{align}
\end{subequations}
If the reaction does not involve chemostatted species, $\tilde \sigma^{(\alpha)}_{\rightarrow,j} =\tilde \sigma^{(\alpha)}_{\leftarrow,j} =0$, we have $\tilde\mu^{(\alpha)}_\rightarrow = \tilde\mu^{(\alpha)}_\leftarrow = 0$, and the flux reduces to the passive one given by \Eqref{eqn:reaction_fluxes_passive}.
In contrast the energy $\Delta \mu^{(\alpha)} = \tilde\mu_\rightarrow^{(\alpha)} - \tilde\mu_\leftarrow^{(\alpha)}$ supplied by the fuel can drive the reaction against the thermodynamic tendency.
The particles exchanged with the chemostat then imply an energy flux across the systems boundary, which corresponds to the entropy production rate in the system if it reaches a stationary state.

\subsection{Phase separation}

Phase separation refers to spontaneous demixing, so we now need to examine inhomogeneous systems.

\subsubsection{Phase equilibrium}
We start by considering a system with several coexisting phases, which are regions of homogeneous composition~$\cs[n]$ where $n$ enumerates the phases.
If  contributions of  interfaces are negligible, the total free energy of this system is given by $F=\sum_n V_n f(\cs[n])$ where $V_n$ is the volume of each phase.
This free energy is minimal when the coexistence conditions,
\begin{subequations}
\label{eqn:coexistence_many}
\begin{align}
	\mu_i \bigl(\cs[1]\bigr) &= \mu_i \bigl(\cs[2]\bigr) = \mu_i \bigl(\cs[3]\bigr) = \cdots
	\label{eqn:coexistence_many_mu}
\\
	\Pi \bigl(\cs[1]\bigr) &= \Pi \bigl(\cs[2]\bigr) = \Pi \bigl(\cs[3]\bigr) = \cdots
	\label{eqn:coexistence_many_P}
	\;,
\end{align}
\end{subequations}
are met for all components~$i$ across all phases~\cite{Weber2019}.
These conditions can be interpreted as chemical and mechanical equilibrium between phases, respectively.
\citet{Gibbs1876} showed that there are at most $N+1$ phases with distinct compositions $\cs^{(1)}, \ldots, \cs^{(N+1)}$ that fulfill \Eqsref{eqn:coexistence_many}.
If we additionally impose incompressibility, only $N$ distinct coexisting phases are possible.

Equilibrium states are conveniently described by partition coefficients~$P^{(nm)}_i= \p[n]_i / \p[m]_i$, which specify the concentration ratio of species~$i$ between different phases $n$ and $m$.
A particularly simple form emerges in incompressible systems when all species have the same molecular volume, so $\sum_i c_i = \mathrm{const}$, leading to
\begin{align}
	\label{eqn:partition_coefficient}
	P^{(n,m)}_i(\cs)
		= \exp\left[
			\enthalpySpec_i(\cs^{(m)}) - \enthalpySpec_i(\cs^{(n)})
		\right]
	 \;,
\end{align}
where we used \Eqsref{eqn:chemical_potential_homogeneous} and \eqref{eqn:coexistence_many}.
This expression reduce to $P_i^{(n,m)} = 1$ in ideal solutions, indicating that phase separation is impossible in such systems.

\subsubsection{Phase separation kinetics}

Without chemical reactions, individual chemical species $X_i$ are conserved and can merely move.
These dynamics can lead to complex spatial patterns, described by concentration fields $c_i(\vect r)$.
Particle conservation implies the continuity equation $\partial_t c_i + \nabla \cdot \vect{j}_i = 0$, where $\vect{j}_i$ denotes the spatial flux of species~$i$.

Similar to the reaction fluxes discussed above, diffusive fluxes generally follow from kinetic models.
However, in contrast to  reactions, a reasonable approximation follows from an expansion around equilibrium states using the framework of linear non-equilibrium thermodynamics, $\vect{j}_i = -\sum_j \Lambda_{ij} \nabla \mu_j$,
where $\Lambda_{ij}$ denotes diffusive mobilities that form a symmetric Onsager matrix~\cite{Julicher2018}.
This illustrates that diffusive fluxes~$\vect{j}_i$ are driven by chemical potential gradients, similar to how reaction fluxes~$s^{(\alpha)}$ are driven by chemical potential differences.
To obtain well-defined equations, the chemical potentials given in \Eqref{eqn:chemical_potential_homogeneous} need to be extended by spatial couplings~\cite{Cahn1958},
\begin{align}
	\label{eqn:chemical_potential_many}
	\mu_i = \kBT \left[
		 \ln\left(\frac{c_i}{\sum_j c_j}\right) + \enthalpySpec_i(\cs)
	   \right] - \sum_{j=1}^N \kappa_{ij} \nabla^2 c_j
	\;,
\end{align} 
where $\kappa_{ij}$ is a coefficient matrix that is related to interfacial tensions~\cite{Weber2019, Mao2018}.
For simplicity, we here skipped the momentum conservation leading to a Navier-Stokes equation~\cite{Anderson1998} and thermal fluctuations~\cite{Cook1970}, which could be added to the kinetic equations~\cite{Julicher2018}.

\subsection{Chemical reactions and phase separation}
We now combine the chemical reaction fluxes, given by \Eqref{eqn:reaction_flux_species}, with the phase separation kinetics described above.
Since species are no longer conserved, the continuity equation becomes $\partial_t c_i + \nabla \cdot \vect{j}_i = s_i$, with the familiar expressions for diffusive fluxes $\vect{j}_i$ and reactive fluxes~$s_i$.
Taken together,
\begin{subequations}
\label{eqn:full_system}
\begin{align}
	\label{eqn:pdes}
	\partial_t c_i &=
		\nabla \cdot \Biggl[ \, \sum_{j=1}^N \Lambda_{ij} \nabla \mu_j \Biggr]
		+ s_i(\cs, \mus)
\\[3pt]
	\label{eqn:reaction_rates}
	s_i(\cs, \mus) &= \sum_{\alpha=1}^M\sigmaA_i k_\alpha
	 \!\left[
		e^{\frac{\tilde\mu^{(\alpha)}_\rightarrow + \sum_j \sigmaA_{\rightarrow,j} \,\mu_j}{\kBT}}
		- e^{\frac{\tilde\mu^{(\alpha)}_\leftarrow + \sum_j \sigmaA_{\leftarrow,j} \,\mu_j}{\kBT}}
	\right]
	\;,
\end{align}
\end{subequations}
where the chemical potentials $\mus=(\mu_1, \ldots, \mu_N)$ are given by \Eqref{eqn:chemical_potential_many}.
These partial differential equations need to be supplemented with boundary conditions.
A typical choice are no-flux conditions, $\vect{j}_i=0$, together with neutral interaction with the system's boundary, implying that the normal derivative of the fields~$\phi_i$ vanish~\cite{Qian2003}.
These equations possess many parameters leading to immense complexity.
Solution strategies range from numerical evolution for specific examples~\cite{Mao2018,Shrinivas2021,Zhou2021a} to full analytical treatment, e.g., via linear stability analysis~\cite{Carati1997}.
We will pursue both approaches for specific examples highlighting the rich behavior.

In a closed, passive systems ($\tilde\mu^{(\alpha)}_\rightarrow=\tilde\mu^{(\alpha)}_\leftarrow=0$), the dynamics described by \Eqref{eqn:full_system} will relax to an equilibrium state.
In such a state, the previously derived conditions for the partition coefficients, \Eqref{eqn:partition_coefficient}, and for the equilibrium ratios of the reactions,  \Eqref{eqn:reaction_equilibrium_ratio}, need to be fulfilled simultaneously.
Combining these two conditions, we obtain
\begin{align}
	\prod_{i=1}^N\left(P^{(n,m)}_i\right)^{\sigmaA_i}
	&= \frac{K^{(n)}_\alpha}{K^{(m)}_\alpha}
	\;,
	\label{eqn:relationship_general}
\end{align}
where $K^{(n)}_\alpha$ denotes the equilibrium ratio of reaction~$\alpha$ in phase~$n$.
This relationship holds for all reactions~$\alpha$ and all pairs of phases $n$ and $m$, which is also evident from the definition of the involved quantities.
The special form given in \Eqref{eqn:relationship_simple} can be recovered for $n=1$, $m=2$, and $\sigma_i = (-1, 1)$ describing the conversion reaction $A \rightleftharpoons B$.
Another special case are components that do not partition, $P^{(n,m)}_i=1$, where \Eqref{eqn:relationship_general} implies that the equilibrium ratios are equal in the compartments, $K^{(n)}_\alpha=K^{(m)}_\alpha$, which is trivially the case in homogeneous systems.

\section{Binary fluid with a conversion reaction}
\label{sec:binary}

\begin{figure}
	\centering	
	\includegraphics{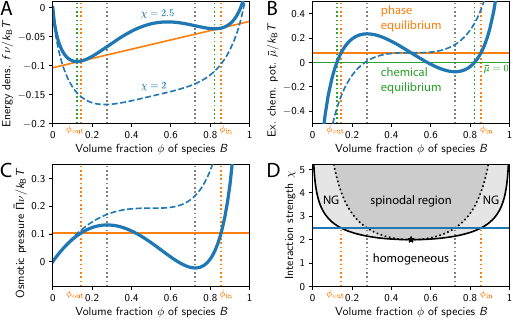}
	\caption{
	\textbf{Equilibrium states of binary systems for different conservation laws.}
	\textbf{(A)} Free energy density $f$ given by \Eqref{eqn:free_energy_binary} as a function of the volume fraction~$\phi$ of species~$B$ for $h(\phi) = w\phi + \chi \phi(1-\phi)$, $w=0.08$, and two interaction parameters~$\chi$.
	The common-tangent construction (orange solid line) determines the equilibrium fractions~$\phiIn$ and $\phiOut$ for $\chi=2.5$ while phase separation is impossible for the critical value $\chi=2$.
	\textbf{(B)} Exchange chemical potential~$\bar\mu$ associated with (A) given by \Eqref{eqn:chemical_potential_binary}.
	Phase equilibrium ($\bar\mu=w$, orange lines) and chemical equilibrium ($\bar\mu=0$, green lines) are indicated.
	\textbf{(C)} Osmotic pressure $\bar\Pi$ associated with (A) given by \Eqref{eqn:pressure_binary}.
	Phase equilibrium is indicated by orange lines.
	\textbf{(D)} Phase diagram as a function of $\phi$ and $\chi$.
	Phase separation is possible above the binodal (solid black line) and homogeneous states are unstable above the spinodal (dotted black line). The two lines enclose the nucleation and growth region (NG) and touch in the critical point (star).
	Colored lines indicate phase equilibrium of previous panels.
	\textbf{(A--D)}
	Homogeneous states are unstable between inflection points of $f(\phi)$ (spinodal, gray dotted lines), while chemical equilibrium is achieved in local minima of $f(\phi)$ (green dotted lines).
	}
	\label{fig:binary_equilibrium}
\end{figure}

To illustrate the behavior of chemical reactions in phase separating systems, we first study a binary fluid that consist of two components, $A$ and $B$.
For simplicity, we now consider an incompressible system, where the molecular volumes $\nu_A$ and $\nu_B$ of the two components are constant and we introduce volume fractions~$\phi_i = \nu_i c_i$.
Since we consider fluids, the two components must fill all available space, implying $\phi_A + \phi_B = 1$.
The system's state is thus specified by one fraction~$\phi$ alone, and the molecular volumes must be equal, $\nu_i=\nu$, to allow chemical transitions.
The free energy density derived from \Eqref{eqn:free_energy_many} becomes
\begin{equation}
	\label{eqn:free_energy_binary}
	f(\phi) = 
		\frac{\kBT}{\nu} \Bigl[
			\phi \ln(\phi)
			+ (1-\phi) \ln(1 - \phi)
			+ h(\phi)
		\Bigr]
	\;,
\end{equation}
where $h(\phi)$ summarizes the enthalpic contributions.
Since the incompressible system does not allow particle insertion, we can only change the composition by replacing an $A$ particle by a $B$ particle.
The associated change in free energy is the exchange chemical potential $\bar\mu = \mu_B - \mu_A$, which reads
\begin{align}
	\label{eqn:chemical_potential_binary}
	\bar\mu(\phi) = \kBT\Bigl[
		\ln(\phi) - \ln(1 - \phi) + h'(\phi)
	\Bigr]
	- \kappa \nabla^2 \phi
	\;,
\end{align}
where we already included the term proportional to $\kappa$, which penalizes gradients akin to \Eqref{eqn:chemical_potential_many}.
Similarly, changing the system's volume requires adding particles, imlpying the Osmotic pressure
\begin{align}
	\label{eqn:pressure_binary}
	\bar\Pi 
	= \frac{\kBT}{\nu}\Bigl[		 
		\phi h'(\phi) - h(\phi)
		-\ln(1 - \phi)
	\Bigr]
	\;.
\end{align}
The relation between $f$, $\bar\mu$, and $\bar\Pi$ is illustrated in \figref{fig:binary_equilibrium}.
The system's dynamics follow from \Eqsref{eqn:full_system} and read~\cite{Weber2019}
\begin{align}
	\label{eqn:dynamics_binary}
	\partial_t \phi = \nabla \cdot \bigl[ \Lambda(\phi) \nabla \bar\mu\bigr] + s
	\;,
\end{align}
where the reaction flux~$s$ denotes the net conversion of $A$ to $B$, which we specify for passive and active systems below.

\subsection{Equilibrium states}

To gain intuition for the system's behavior, we first discuss various equilibrium states.
Here, we distinguish three different cases: (1) homogeneous systems that allow chemical transitions, $A \rightleftharpoons B$, (2) heterogeneous systems that conserve the individual species $A$ and $B$, and (3) the general case where both chemical transitions and spatial inhomogeneities happen.
Case (2) corresponds to model~B in the classification by \citet{Hohenberg1977}, while the other cases are related to model~A.

\subsubsection{Chemical equilibrium}
The only possible reaction, $A \rightleftharpoons B$, is equilibrated when $\bar\mu = 0$, which corresponds to the minima of the free energy density~$f(\phi)$ (green dotted lines in \figref{fig:binary_equilibrium}).
There can be multiple (local) minima and the height of the energy barrier between them determines the transition rates~\cite{Hanggi1990}.
If we included thermal noise, the system would explore the entire energy landscape according to Boltzmann's distribution.

The equilibrium states can be determined for simple systems.
For instance, if particles do not interact, $h(\phi)=h_0 + w\phi$, the equilibrium ratio~$K=\phi_B/\phi_A$ only depends on the internal energy difference~$w$ between $B$ and $A$,
\begin{align}
	K &= e^{-w}
	\;, 
\end{align}
which follows from \Eqref{eqn:chemical_potential_binary}.
The state contains less $B$ (smaller $K$) when its internal energy is larger than that of $A$ ($w>0$).

The simplest interaction between $A$ and $B$ is captured by a Flory-parameter~$\chi$,  $h(\phi) = h_0 + w \phi + \chi \phi(1 - \phi)$~\cite{Flory1942}.
When $\chi$ is sufficiently large ($\chi > 2$ for $w=0$), the free energy exhibits a local maximum, which separates two minima that both correspond to reaction equilibria; see \figref{fig:binary_equilibrium}.
The relative energies of these state, and thus the transitions in this bistable system, can be independently controlled by the parameter~$w$.

\subsubsection{Phase separation equilibrium}
Phase separation is impossible in ideal, non-interacting system.
We thus discuss the non-ideal system described by $h(\phi) = h_0 + w \phi + \chi \phi(1 - \phi)$, but we now switch off chemical reactions.
Considering an inhomogeneous system with two phases of compositions $\phi^{(1)}$ and $\phi^{(2)}$ that can exchange particles and volume, the coexistence conditions~\eqref{eqn:coexistence_many} become
\begin{subequations}
\label{eqn:coexistence_binary}
\begin{align}
	0 &= f'(\phi^{(1)}) - f'(\phi^{(2)})
\\
	0 &= f(\phi^{(1)})  - f(\phi^{(2)}) + f'(\phi^{(1)}) \bigl( \phi^{(2)} -  \phi^{(1)}\bigr)
	\;.
\end{align}
\end{subequations}
These equations can be solved graphically by a common tangent construction, or \emph{Maxwell construction}; see orange line in \figref{fig:binary_equilibrium}A.
The tangent points $\phi^{(1)}$ and $\phi^{(2)}$ denote the fractions in the coexisting phases, while the common tangent marks the average free energy density of the phase separated system with average volume fraction~$\phi$.
This construction thus shows that the free energy of the system can be reduced by splitting into two phases of 
composition $\phi^{(1)}$ and $\phi^{(2)}$ when $f(\phi)$ has a concave region, i.e., when $f''(\phi) < 0$.
For the particular $h(\phi)$ discussed here, this \emph{spinodal region} is given by 
\begin{align}
	\label{eqn:binary_spinodal}
	\frac{1}{\chi +\sqrt{(\chi -2) \chi }} < \phi < \frac{1}{\chi -\sqrt{(\chi -2) \chi}}
	\;,
\end{align}
which is only valid for $\chi>2$, which marks the critical interaction strength; see dark gray region in \figref{fig:binary_equilibrium}D.
Homogeneous states with fractions that satisfy this condition are unstable and phase separate immediately.
Conversely, one can show that phase separated states are unstable outside the \emph{binodal region}, which is delineated by the equilibrium fractions $\phi^{(1)}$ and $\phi^{(2)}$ as a function of $\chi$; see light gray region in \figref{fig:binary_equilibrium}D.
In the light gray region between the binodal and the spinodal, both the homogeneous and the phase separated state are (meta)stable, so that droplets only form after stochastic nucleation~\cite{Cahn1959,Binder1976a,Hoyt1990,Xu2014}.

\subsubsection{Combined equilibrium}
\label{sec:binary_combined_equilibrium}
If phase separation and chemical reactions can take place together, the binary system only exhibits homogeneous equilibrium states.
This is because the chemical reaction does not conserve  individual particle counts and the system can thus attain a composition that minimizes $f(\phi)$ everywhere.
Consequently, the combined equilibrium of a binary system is completely dominated by chemical reactions.

\subsection{Relaxation toward equilibrium}
\label{sec:binary_relaxation}
\Eqref{eqn:dynamics_binary} describes how an arbitrary initial state relaxes toward the equilibrium states.
This equation assumed linear non-equilibrium thermodynamics for the spatial (diffusive) fluxes and we here will employ the same approximations for the reactions.
Linearizing the flux~$s$ following from the transition state theory given in \Eqref{eqn:reaction_rates} for small $\bar\mu$, we obtain the passive reaction flux $\sPa = -\mobP \bar\mu$.
Here, $\mobP = k/\kBT$ is the reaction mobility, which must be positive to obey the second law.
The diffusive mobility~$\Lambda$ is typically chosen as $\Lambda(\phi) = \Lambda_0 \phi(1-\phi)$~\cite{Kramer1984}, so the dynamics reduce to ordinary diffusion with diffusivity $D= \kBT \Lambda_0$ in the dilute limit ($\phi \ll 1$).

\begin{figure}
	\centering	
	\includegraphics[width=252pt]{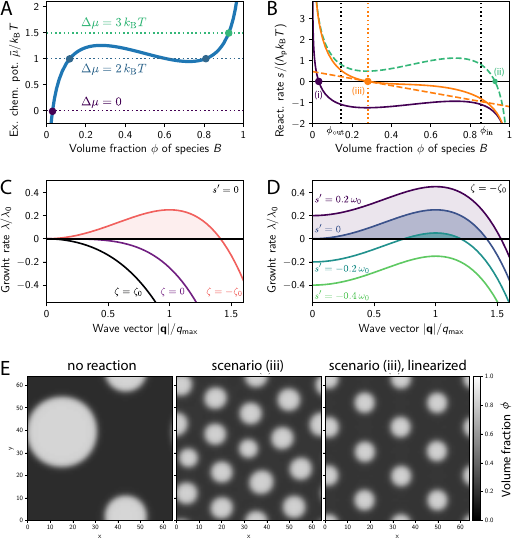}
	\caption{
	\textbf{Binary active fluids can form regular patterns.}
	\textbf{(A)} The exchange chemical potential~$\bar\mu$ as a function of composition~$\phi$.
	Reactions cease when their chemical potential difference obeys \Eqref{eqn:stationary_state_active_homogeneous}, which depends on the external energy~$\Delta\mu$.
	Here, $\mobA = \mobP$.
	\textbf{(B)} Reaction flux $s$ as a function of $\phi$ for various reaction models:
	(i) Passive reactions ($\mobA=0$),
	(ii) Active reactions with constant mobilities $\mobA = \mobP$, and
	(iii) Active reactions with mobility $\mobA(\phi) = \mobP (1 - \phi)$ proportional to $\phi_A=1-\phi$.
	The orange dashed line denotes the linear expansion around the stationary point of case (iii).
	\textbf{(C, D)} Growth rate~$\lambda$ of a perturbation with wave vector~$\vect{q}$ for various parameters $\zeta$ and $s'$; see \Eqref{eqn:binary_linear_stability}.
	Quantities are normalized to $q_\mathrm{max} = (-\zeta_0/[2 \kappa  \Lambda(\phi_*)])^{1/2}$ and $\lambda_0 = \zeta_0^2/[4\kappa\Lambda(\phi_*)]$ for $\zeta_0>0$.
	\textbf{(E)} Numerical simulations of no reaction ($\mobA=\mobP=0$) and the two versions of scenario (iii) from panel B.
	Simulations were performed using finite differences~\cite{Zwicker2020} with periodic boundary conditions and non-dimensional units ($\Lambda=\kBT=\nu=1$) for $10^4$ time units and $\mobP=1$.
	\textbf{(A, B, E)} Additional parameters are $h(\phi) = w\phi + \chi \phi(1-\phi)$ with $w=1.1$, $\chi=2.5$, and $\Delta\mu = 3\,\kBT$.
	}
	\label{fig:binary_active}
\end{figure}

The relaxation dynamics can be analyzed in limiting cases.
If only chemical reactions ($\Lambda_0=0$) or only phase separation ($\mobP=0$) take place, these equations reduce to the classical models A and B, respectively~\cite{Hohenberg1977}.
In both cases, phase separation with complex structures of many droplets or bicontinuous structures may develop initially.
In the case of chemical reactions (non-conserved, model A), the characteristic length scales grow as $t^{1/2}$ with time~$t$ until the system reaches a homogeneous state in equilibrium~\cite{Bray2003}.
In contrast, in the  case of phase separation (conserved, model B), the length scale grows as $t^{1/3}$, according to the theory by Lifshitz, Slyozov, and Wagner~\cite{Lifshitz1961,Wagner1961}, until a single droplet remains in equilibrium.
The mixed case, where both chemical reactions and phase separation take place, is more difficult to analyze since both scaling regimes may contribute to the dynamics.
However, the scaling $t^{1/2}$ of model A will dominate after long times, and the system ends up in the homogeneous equilibrium state.

\subsection{Active systems}
\label{sec:binary_active}

While chemical reactions limit passive, binary systems to homogeneous equilibrium states, active systems may display richer behaviors.
For binary systems, the only possible active reaction converts $A$ and $B$ into each other using external energy~$\Delta\mu$ supplied by some fuel.
Adding the driven reaction flux described by \Eqref{eqn:reaction_fluxes_active}, the total reaction flux reads $s = \sPa + \sAc$,
where we again consider a linearized reaction flux, $\sAc= - \mobA (\phi) (\bar\mu - \Delta\mu)$.
Interesting dynamics emerge when the mobility~$\mobA$ depends on composition~$\phi$.

\subsubsection{Homogeneous systems}
To unveil effects of the active reaction, we first discuss discuss homogeneous systems.
Stationary states ($s=0$) obey
\begin{align}
	\label{eqn:stationary_state_active_homogeneous}
	\bar\mu_* = \frac{\mobA\Delta\mu}{\mobP + \mobA}
	\;,
\end{align}
which reduces to chemical equilibrium ($\bar\mu_*=0$) in the passive case ($\Delta\mu=0$).
In any other case, the stationary state is not in thermodynamic equilibrium.
This can be seen by looking at the reaction fluxes, which read $\sAc_* = -\sPa_* =\Delta\mu \mobA\mobP/(\mobA + \mobP)$.
While the total flux $\sPa_* + \sAc_*$ indeed vanishes, both chemical reactions  continuously convert the two species into each other.
The energy invested in this system is given by the density of the entropy production rate, $\sAc_* \Delta\mu$, which is always positive and increases with stronger drive.
Since the invested energy does not result in any apparent interesting dynamics, such reactions are also called \emph{futile cycles}.
However, they have been shown to produce oscillatory behavior in stochastic systems~\cite{Samoilov2005}.
Moreover, both the energy input~$\Delta\mu$ and the reaction mobilities, which could be affected by enzymes, influence the stationary value of $\bar\mu$; see \Eqref{eqn:stationary_state_active_homogeneous}.
Consequently, enzyme activity  can control the associated equilibrium ratio~$K= \phi_*/(1-\phi_*)$ and the actually attained stationary state in the bistable situation (see \figref{fig:binary_active}A), i.e., whether the system settles into a state enriched in $A$ or $B$.

\subsubsection{Heterogeneous systems}
So far, we discussed homogeneous systems, but the dynamics described by \Eqref{eqn:dynamics_binary} also permit phase separation.
In particular, stationary homogeneous state might no longer be stable and instead develop heterogeneities.
To elucidate this, we perform a linear stability analysis by considering the perturbation $\phi(\vect r, t) = \phi_* + \epsilon \exp(\lambda t + \iu \vect q \cdot \vect r)$.
Expanding \Eqref{eqn:dynamics_binary} to linear order in $\epsilon$, we find~\cite{Weber2019}
\begin{align}
	\label{eqn:binary_linear_stability}
	\lambda(\vect q) &= s'(\phi_*) - \vect{q}^2 \zeta(\phi_*) - \vect{q}^4 \Lambda(\phi_*) \kappa
\end{align}
where $\zeta(\phi_*) = \Lambda(\phi_*) f''(\phi_*) + \kappa (\mobA + \mobP)$.
The growth rate~$\lambda$ can be positive, and the homogeneous state thus unstable, if $\zeta<0$, implying $f''(\phi_*)<0$; see \figref{fig:binary_active}C.
For systems without reactions ($\mobP=\mobA=0$), this corresponds to the spinodal region and is consistent with coarsening since arbitrarily low wave vectors (corresponding to long wave lengths) are unstable.
In contrast, equilibrium states with passive chemical reactions ($\mobA=0$, $\mobP>0$) are characterized by a minimal free energy density, implying $f''(\phi_*) > 0$.
Finally, active chemical reactions allow more diverse behavior since $\zeta$ and $s'(\phi_*)$ can in principle be controlled separately.
In particular, a band of unstable wave vectors may emerge when $\zeta<0$ and $s'<0$; see \figref{fig:binary_active}D.
This suggests that homogeneous states are unstable while coarsening to arbitrarily large wave lengths is also suppressed.
The corresponding states must thus be quite different to what we have discussed so far.

\subsubsection{Active droplets}
We have seen that constant reaction mobilities $\mobP$ and $\mobA$ imply homogeneous states, while more complex reactions can destabilize such states.
To discuss a concrete situation, we consider $\mobA = \mobP(1-\phi)$ for constant~$\mobP$, so that the active reaction is suppressed for large~$\phi$.
\figref{fig:binary_active}E shows that this choice leads to a regular pattern of droplets of similar size, which we term \emph{active droplets} since their size is controlled by the driven chemical reactions.

To understand why the driven reaction with concentration-dependent mobility can control the size of active droplets, we further simplify the system.
\figref{fig:binary_active}B shows that the reaction rate~$s =  \sPa + \sAc$ decreases monotonically with~$\phi$ and has a reaction equilibrium slightly above the equilibrium fraction~$\phiOut$ in the dilute phase.
Starting with a dilute phase in phase equilibrium, the chemical reactions thus constantly produce droplet material~$B$, which can nucleate new droplets or fuel the growth of existing ones.
However, the fraction inside such droplets is so large that the reactions effectively convert droplet material~$B$ to pre-cursor~$A$; $s(\phiIn) < 0$.
These pre-cursors leave the droplet to join the dilute phase, thus closing a cycle driven by dissipating energy.
This situation is qualitatively captured by using a linear reaction rate, $s=-k(\phi-\phi_*)$, where $\phi_*$ is the reaction equilibrium, which must obey $\phiOut < \phi_* < \phiIn$, and $k$ sets the time scale of the dynamics; see dashed orange line in \figref{fig:binary_active}B.
The linearized dynamics yield qualitatively similar behavior (see \figref{fig:binary_active}E) and have been studied before~\cite{Glotzer1994,Glotzer1994a,Toxvaerd1996,Motoyama1996,Zwicker2015} although the thermodynamic details were debated~\cite{Lefever1995,Glotzer1995a}.
Interestingly, the stationary state of the system with the linearized active reaction can be formally rewritten as an equilibrium state of another system where the reaction has been replaced by long-ranged interactions~\cite{Liu1989}.
This analogy allows exploiting equilibrium thermodynamics to study the properties of this inherently non-equilibrium system~\cite{Sagui1995,Christensen1996,Motoyama1997,Muratov2002}.

To finally see how droplets of a particular size emerge from driven reactions, we focus on a single, isolated droplet~\cite{Zwicker2015}.
The droplet volume~$V$ changes due to material influx~$J$ from the surrounding dilute phase and due to the integrated reaction flux~$S$ inside the droplet that removes droplet material, $\partial_t V \approx (J-S)/\phiIn$~\cite{Weber2019}.
In the simplest case, the influx is diffusion-limited, implying $J \approx 4\pi D R \epsilon$, where $\epsilon= \phi_* - \phiOut$ is the supersaturation created by the chemical reaction~\cite{Weber2019}.
Conversely, the reaction inside the droplet approximately implies $S \approx k(\phiIn-\phi_*)V$.
Taken together, this implies a stationary droplet radius, 
\begin{align}
	R_* &\approx \left[\frac{3D (\phi_* - \phiOut)}{k(\phiIn-\phi_*)}\right]^{\frac12}
	\;.
\end{align}
The mean distance~$L_*$ between droplets follows from the fraction~$\enthalpySpec$ of the system occupied by droplets.
For linear reactions, integrating \Eqref{eqn:dynamics_binary} over the system's volume and using no-flux boundary conditions implies that the average fraction equals~$\phi_*$.
We thus find $\enthalpySpec = (\phi_* - \phiOut)/(\phiIn - \phiOut)$ and $L_* \propto \enthalpySpec^{1/3} R_*$.
Consequently, the length scales $R_*$ and $L_*$ are both governed by the reaction-diffusion length scale $\sqrt{D/k}$, akin to traditional Turing patterns~\cite{Turing1952}.
Note that these deceivingly simple results required several approximations, which might not always be warranted~\cite{Kirschbaum2021}.
More importantly, the conversion $A \rightleftharpoons B$ and the phase separation of the components are intimately linked in this simple case of binary systems.
Experimentally relevant examples typically contain more species, where these two processes might be more independent.

\begin{figure*}[t]
	\centering	
	\includegraphics[width=522pt]{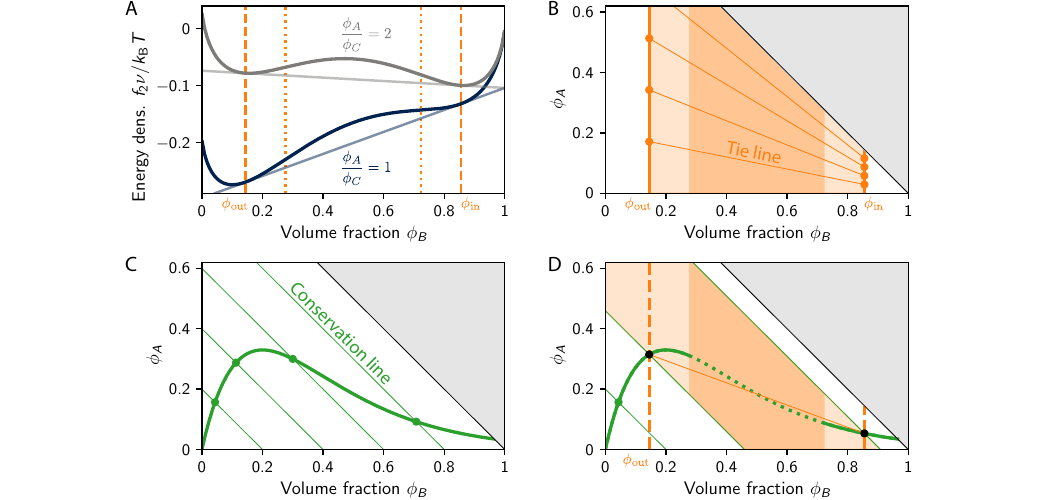}
	\caption{
	\textbf{Equilibrium states of ternary systems for different conservation laws.}
	\textbf{(A)} Free energy density $f$ as a function of $\phi_B$ for two fixed ratios~$\eta=\phi_A/\phi_C$.
	Both tangent constructions (faint lines) yield the same equilibrium fractions~$\phiIn$ and $\phiOut$ (dashed orange lines) and spinodal points (dotted orange lines).
	\textbf{(B)} Phase equilibrium (no reactions) implies that phase separation is possible in the light orange region between the equilibrium fractions~$\phiIn$ and $\phiOut$.
	The tie lines (thin lines) indicate how an initially homogeneous system separates into two coexisting equilibrium phases (dots).
	Homogeneous states are unstable in the dark orange spinodal region.
	\textbf{(C)} Chemical equilibrium of a homogeneous system is obeyed on the thick line. The conservation law $\psi=\phi_A + \phi_B$ is visualized as thin conservation lines and equilibria are indicated by dots.
	\textbf{(D)} In the combined equilibrium, there are two possible equilibrium states:
	Homogeneous states are governed by chemical equilibrium (green solid line), which is now only stable outside the spinodal region (dark orange area).
	Conversely, phase separated states can emerge for initial conditions in the orange shaded regions, where chemical equilibrium selects a particular tie line (thin orange line) along which the system phase separates into coexisting phases (black dots).
	\textbf{(A--D)}
	Parameters are $w_A=1$, $w_B=0$, and $\chi=2.5$.
	The gray region is forbidden since $\phi_A + \phi_B>1$. 
	Plots were inspired by \cite{Bauermann2021}.
	}
	\label{fig:ternary_equilibrium}
\end{figure*}

\section{Ternary fluid with a conversion reaction}
\label{sec:ternary}
We next discuss systems with three components, where phase separation and chemical transitions are less intertwined than in binary systems.
This allows for richer behavior, thus giving a better picture of real systems.
In fact, even equilibrium phase separation is already complex in ternary systems since now many phases with different compositions may form~\cite{Gonzalez-Leon2003, Mao2018, Devirie2021}.

To limit complexity, we consider a specific example where two components $A$ and $B$ are interconverted, $A \rightleftharpoons B$, and the third component~$C$ plays the role of an inert solvent.
The system's state is given by the two fractions~$\phi_A$ and $\phi_B$, while the solvent fraction is $\phi_C=1-\phi_A-\phi_B$.
Phase separation is driven by an effective repulsive interaction between $B$ and the other components, which is captured by
$\enthalpy(\phi_A, \phi_B) = w_A \phi_A + w_B \phi_B + \chi \phi_B(1 - \phi_B)$,
where $w_A$ and $w_B$ are internal energies and $\chi$ is a Flory-parameter that governs interactions.
The associated exchange chemical potentials read
\begin{subequations}
\label{eqn:ternary_chemical_potentials}
\begin{align}
	\label{eqn:ternary_muA}
	\bar\mu_A &= \kBT\Bigl[
		\ln (\phi_A) - \ln (\phi_C) + w_A
	\Bigr]
\\
	\label{eqn:ternary_muB}
	\bar\mu_B &= \kBT\Bigl[
		 \ln (\phi_B) - \ln (\phi_C) + w_B + \chi(1 - 2\phi_B)
	\Bigr]
\end{align}
\end{subequations}
where $\phi_C = 1 - \phi_A - \phi_B$.
We will see that these chemical potentials not only govern the equilibrium states, but also inform the behavior of active systems.

\subsection{Equilibrium states}

We start by analyzing phase equilibrium without chemical reactions, so amounts of the three species $A$, $B$, and $C$ are conserved individually.
For the specific system given by \Eqref{eqn:ternary_chemical_potentials},  phase equilibrium implies constant ratios of the fractions of $A$ and $C$ in all phases, $\phi_A^{(1)}/\phi_C^{(1)} = \phi_A^{(2)}/\phi_C^{(2)} = \eta$, which is conserved and thus set by the initial condition.
The free energy then reduces to the binary form given in \Eqref{eqn:free_energy_binary} with $h(\phi_B) = \chi\phi_B(1-\phi_B) + w\phi_B + h_0$, where $w$ and $h_0$ only depend on the constants $w_A$, $w_B$, and $\eta$.
Consequently, phase separation is possible for~$\chi>2$, the spinodal region is given by \Eqref{eqn:binary_spinodal}, and the equilibrium fractions $\phiOut$ and $\phiIn$ are independent of $\eta$; see \figref{fig:ternary_equilibrium}A--B.
The remaining fraction $1-\phi_B$ is occupied by $A$ and $C$ in the pre-determined ratio~$\eta$ in each phase. %
The special choice of the free energy thus fully decouples phase separation of $B$ from dynamics that determine the ratio of $A$ to $C$.

If we now allow the chemical transition $A \rightleftharpoons B$, only the total amount of $A$ and $B$ is conserved, while their ratio may vary.
The conserved quantity is thus $\psi = \bar\phi_A + \bar\phi_B$, where $\bar\phi_i = \Vsys^{-1} \int \phi_i \, \diff V$ is the average fraction of species~$i$ in the system.
Since the solvent~$C$ does not participate in the reaction, its amount is also conserved, $\bar\phi_C = 1- \psi$.
Conversely, the equilibrium ratio between $A$ and $B$, quantified by the ratio $K=\phi_B/\phi_A$, is governed by chemical equilibrium ($\bar\mu_A = \bar\mu_B$) and reads
\begin{align}
	\label{eqn:ternary_reaction_equilibrium}
	K = e^{w_A - w_B - \chi(1 - 2\phi_B)}
	\;,
\end{align}
which follows from \Eqref{eqn:ternary_chemical_potentials}.
This condition traces out a line in the phase space spanned by $\phi_A$ and $\phi_B$; see \figref{fig:ternary_equilibrium}C.
Actual equilibrium states are then determined by conservation lines (where $\phi_A + \phi_B = \psi$) set by  initial conditions.

\begin{figure}
	\centering	
	\includegraphics[width=252pt]{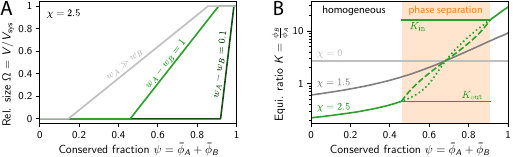}
	\caption{
	\textbf{Regulation of phase behavior in passive systems.}
	\textbf{(A)} Volume $V$ of dense phase relative to system volume $\Vsys$ as a function of the conserved fraction~$\psi=\bar\phi_A + \bar\phi_B$ for several internal energy differences $w_A - w_B$ for $\chi=2.5$.
	\textbf{(B)} Equilibrium ratio $K=\phi_B/\phi_A$ as a function of $\psi$ for several interaction parameters~$\chi$ for $w_A-w_B=1$.
	The system is generally homogenous, apart from the case $\chi=2.5$ in the orange shaded region (compare to \figref{fig:ternary_equilibrium}D).
	In this case, there are two phases with different ratios $K_\mathrm{in}$ and $K_\mathrm{out}$ and the ratio averaged over the entire system ($\mean{K}$, dashed line) deviates from the value predicted for the (unstable) homogeneous state (dotted line).
	}
	\label{fig:ternary_regulation}
\end{figure}

Phase equilibrium and chemical equilibrium each select equilibrium states based on equilibrium conditions and conservation laws, which are respectively visualized as thick and thin lines in \figref{fig:ternary_equilibrium}B--C.
In contrast, when phase separation and chemical transitions take place, both equilibrium conditions must be satisfied simultaneously, while neither of the conservation laws hold.
Consequently, intersections of the lines marking the equilibrium conditions denote equilibrium states; see black dots in \figref{fig:ternary_equilibrium}D.
In fact, all phase separated systems initialized within the orange region will converge toward these two points.
There are thus fewer possible equilibrium configurations compared to either phase separation or chemical transitions alone.
This is analogous to the binary case discussed in \secref{sec:binary_combined_equilibrium} where adding a chemical transition to phase separation reduced the complexity of possible equilibrium states tremendously.
We saw that equilibrium states of ternary systems with a chemical transition reduce to phase separation of a binary system without reactions, so it seems as if reactions do not matter.
However, the details of the reaction determine what binary system is selected and thus influence global quantities like the volume~$V$ of the $B$-rich phase.
\figref{fig:ternary_regulation}A shows that $V$ becomes nonzero beyond a threshold value of the conserved fraction~$\psi=\bar\phi_A + \bar\phi_B$.
The threshold is determined by the condition $\bar\phi_B = \phiOut$, which marks the left binodal line in \figref{fig:ternary_equilibrium}B.
For larger~$\psi$, the volume $V$ increases linearly until the right binodal line ($\bar\phi_B = \phiIn$) is reached.
This behavior is expected from binary phase separation, but a crucial difference is that the internal energy difference $w_A - w_B$ now determines the ratio of $A$ to $B$ components and thus the precise transition points.
Consequently, the chemical transition can control phase separation.

Phase separation also affects the chemical transition.
In particular, the equilibrium ratio~$K$, given by \Eqref{eqn:ternary_reaction_equilibrium}, is only a constant in ideal systems ($\chi=0$).
In non-ideal systems with $\chi>0$, the equilibrium ratio increases with increasing fraction~$\psi$, even if the system stays homogeneous; see \figref{fig:ternary_regulation}B.
For $\chi > 2$, the system can separate into two phases of different composition, and thus different equilibrium ratio according to \Eqref{eqn:relationship_general}.
Note that the equilibrium ratio averaged over the entire system, $\mean{K}=\bar\phi_B/\bar\phi_A$, increases continuously with the fraction~$\psi$, albeit with a slightly different functional form than the equilibrium ratio predicted for the (unstable) homogeneous phase; see \figref{fig:ternary_regulation}B.
Consequently, the fact that phase separation takes place might be difficult to detect from the overall reaction balance, but local differences between phases can be substantial.

\subsection{Relaxation toward equilibrium}
The relaxation dynamics are described by  \Eqref{eqn:full_system} using $\tilde\mu^{(\alpha)}_{\rightleftharpoons} = 0$ and chemical potentials defined in \Eqref{eqn:ternary_chemical_potentials}.
These equations can generally only be solved numerically, but qualitative insight can be drawn from limiting cases.
In particular, the limiting cases of exclusive chemical reactions or phase separation will behave similarly to the binary system discussed in \secref{sec:binary_relaxation}.
The more complex case of combined phase separation and chemical reactions is less well explored and we will briefly discuss limiting cases below.

If chemical transitions are fast compare to diffusive fluxes, any initial condition quickly relaxes to chemical equilibrium locally before spatial fluxes become important.
Graphically, this corresponds to a projection of the full phase space along conservation lines onto the line of chemical equilibrium (green line in \figref{fig:ternary_equilibrium}D).
The chemical transition confines the system to this manifold and the spatial dynamics are then those of the binary system, which we discussed in \secref{sec:binary_relaxation}.
Taken together, this limiting case does reproduce ordinary phase separation kinetics, including Ostwald ripening.

The converse case of fast diffusion is analyzed thoroughly by \citet{Bauermann2021}.
In this case, the system quickly phase separates and slaves the fractions in each phase to the binodal line (vertical orange lines in \figref{fig:ternary_equilibrium}B).
The fraction in the different phases then slowly evolves along the binodal line until it reaches the equilibrium state (black dots in \figref{fig:ternary_equilibrium}D).

The dynamics of systems with phase separation and  chemical reactions will strongly depend on the associated mobilities $\mob$ and $\mobP$, as well as the typical length scale $L$ of the fields.
If $\mob \gg \mobP L^2$, diffusive dynamics are fast compared to reactions, leading to the case discussed by \citet{Bauermann2021}.
Conversely, chemical reactions dominate  if $\mob \ll \mobP L^2$.
Since coarsening processes increases $L$, reaction dynamics will always dominate after a long time in a sufficiently large system.

\subsection{Active systems}

We now drive the ternary system out of equilibrium by introducing an active reaction that uses the externally supplied energy~$\Delta\mu$ to drive the transition from $A$ to $B$, similar to the binary case discussed in \secref{sec:binary_active}.
Again using linear approximations of the reaction fluxes, the active and the passive transition balance when
\begin{align}
	\label{eqn:ternary_active_homogeneous}
	\bar\mu_B - \bar\mu_A = \frac{\mobA\Delta\mu}{\mobP + \mobA}
	\;,
\end{align}
akin to \Eqref{eqn:stationary_state_active_homogeneous}.
Here, $\mobA$ and $\mobP$ are the mobilities of the active and passive reaction, respectively.
If both mobilities are constant, the active reaction only shifts the chemical balance between species~$A$ and $B$, which is equivalent to modifying the internal energy difference $w_B-w_A$.
Consequently, the behavior of this system reduces to the passive case discussed above, although stationary states are not in equilibrium, since the chemical energy~$\Delta\mu$ is used continuously to convert the molecules in a futile cycle.

\begin{figure}
	\centering	
	\includegraphics[width=252pt]{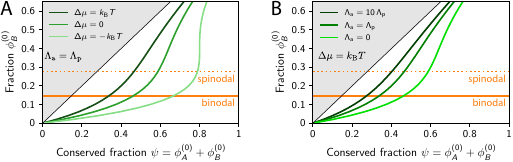}
	\caption{
	\textbf{Regulation of chemical composition in passive systems.}
	Fraction $\phi_B^{(0)}$ in the stationary state obeying \Eqref{eqn:ternary_active_homogeneous} as a function of the conserved fraction $\psi=\phi_A^{(0)} + \phi_B^{(0)}$ in a homogeneous state for 
	(A) several $\Delta\mu$ at $\mobA=\mobP$, and
	(B) several ratios $\mobA/\mobP$ at $\Delta\mu=\kBT$.
	Phase separation is possible above the binodal ($\phi_B^{(0)} = \phiOut$, solid orange line) and it is the only stable state above the spinodal (dotted orange line).
	Model parameters are $\chi=2.5$ and $w_A - w_B = 1$.
	}
	\label{fig:ternary_active_homogeneous}
\end{figure}

To find interesting behavior in active ternary systems, at least one of the mobilities must depend on composition.
In experiments, this generic case can be controlled by introducing enzymes that catalyze the passive or active reaction and by controlling the affinity of the enzymes for either phase.
Similarly, the affinity of the fuel could be controlled, making $\Delta\mu$ dependent on composition~\cite{Bartolucci2021}.
These possibilities lead to many different dependencies of $\mobP$ and $\mobA$ on the composition.
To limit complexity, we will focus on the relevant case where few $B$-rich droplets coexist with a dilute phase that occupies most of the system.
This separation of length scales allows us to first focus on the dilute phase and then treat droplets as a perturbation.

We assume that a dilute phase with a particular composition, given by $\phi_A^{(0)}$ and $\phi_B^{(0)}$, exists.
This state must obey the stationary state condition~\ref{eqn:ternary_active_homogeneous} where the mobilities are evaluated at the composition.
Similarly to the binary case, the mobilities and $\Delta\mu$ influence the stationary state; see 
\figref{fig:ternary_active_homogeneous}.
In particular, the parameters of the chemical reaction, together with the conserved fraction~$\psi=\phi_A^{(0)}+\phi_B^{(0)}$, determine whether the dilute phase is under-saturated ($\phi^{(0)}_B < \phiOut$) or super-saturated ($\phi^{(0)}_B > \phiOut$).
Since these two cases have very different dynamics~\cite{Soeding2019}, we discuss them separately.

\subsubsection{Externally maintained droplets}
\begin{figure}[t]
	\centering	
	\includegraphics[width=252pt]{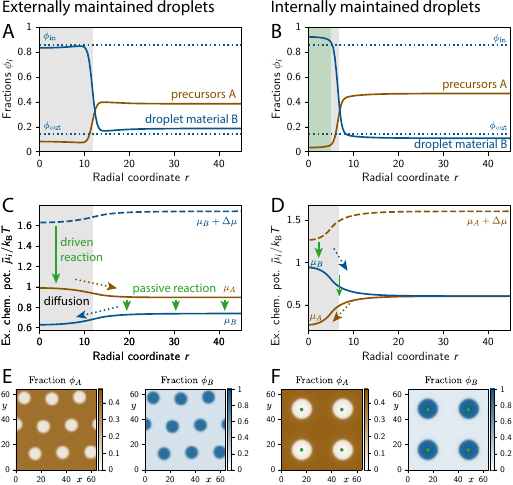}
	\caption{
	\textbf{Behavior of active droplets in a ternary fluid.}
	Droplet material is produced outside droplets in externally maintained droplets (left column), while a localized reaction produces droplet material inside internally maintained droplets (right column).
	\textbf{(A, B)} Volume fractions~$\phi_i$ as a function of the radial coordinate~$r$ of a spherically symmetric system.
	Droplets are regions where $\phi_B > \frac12$ (gray area).
	The green region in B indicates the localized production.
	\textbf{(C, D)} Exchange chemical potentials~$\bar\mu_i$ as a function of $r$.
	Gradients in $\bar\mu_i$ induce diffusive fluxes (dotted arrows), while chemical potential differences drive chemical reactions (green arrow).
	The driven reaction incorporates the external energy~$\Delta\mu$ to force reaction against the passive tendency, which drives cyclic fluxes in the system.
	\textbf{(E, F)} Fractions~$\phi_A$ and $\phi_B$ in stationary state.
	\textbf{(A--F)}
	All panels show stationary states of \Eqref{eqn:full_system} using \Eqref{eqn:ternary_chemical_potentials}, which were obtained using finite-differences~\cite{Zwicker2020} in a 3D spherical geometry (A--D) and a 2D Cartesian geometry with periodic boundary conditions (E, F).
	Model parameters in non-dimensional units ($\Lambda=\kBT=\nu=1$) are
	$\chi=2.5$, $w_B=0$, $\kappa=0.1$, $\Delta\mu=1$, and $\mobP=0.01$.
	For externally maintained droplets $w_A=1$ and $\mobA = \phi_B \mobP$,
	while for internally maintained droplets $w_A=0.5$ and the mobility is localized: 
	$\mobA = 0.1$ for $r<5$ in (B, D) and at the green spots in (F).
	}
	\label{fig:ternary_active_droplets}
\end{figure}

If the dilute phase is super-saturated ($\phi^{(0)}_B > \phiOut$), droplets form spontaneously, either by nucleation or via spinodal decomposition.
Droplet growth is limited either by depleting the dilute phase or by additional reactions taking place inside the droplet.
In the first case, $A$ and $B$ components are transferred from the dilute phase to the droplets until the fraction $\psi$ in the dilute phase decreases such that $\phi^{(0)}_B=\phiOut$; see \figref{fig:ternary_active_homogeneous}A.
At this point, the diffusive influx ceases and droplet growth stops.
In this case, the active reactions merely controls the balance between $A$ and $B$ inside and outside the droplet, but the stationary state still contains homogeneous phases, and droplets will exhibit Ostwald ripening~\cite{Kirschbaum2021}.
The second case of an additional chemical reaction is more interesting:
If the chemical mobilities~$\mobP$ and $\mobA$ are set up such that the reactions effectively destroy droplet material inside the droplet, $s(\phiIn) < 0$, a continuous cycle is maintained:
Droplet material~$B$ is converted to precursor~$A$ inside the droplet, which then leaves the droplet toward the dilute phase, where it is converted back to $B$ and joins a droplet.
The left column of \figref{fig:ternary_active_droplets} shows a concrete example of such a system, where the cyclic fluxes are particularly visible in the chemical potentials shown in panel C.
Since droplet material is created outside of droplets, we call these systems \emph{externally maintained droplets}~\cite{Weber2019}.
The magnitude of the diffusive fluxes depend on the droplet size, so that typically a stable droplet size emerges~\cite{Kirschbaum2021}; see \figref{fig:ternary_active_droplets}E.
Note that the opposite case where droplet material is produced inside the droplet, $s(\phiIn) > 0$, implies that all phases produce droplet material, which will lead to a single homogeneous stationary state enriched in droplet material.
Taken together, we thus demonstrated how carefully regulated chemical reactions can promote droplet growth by a super-saturated dilute phase, while they also limit droplet growth via degradation of droplet material inside droplets.
This principle can also lead to shape instabilities of droplets, which then divide spontaneously~\cite{Zwicker2017, Seyboldt2018}.

The simplest realization of externally maintained droplets involves first-order chemical reactions, $s(\phi) = -k(\phi-\phi^{(0)})$, which clearly display the controlled droplet size~\cite{Zwicker2015}.
However, these first-order reactions are difficult to reconcile with the thermodynamic principles that we introduced in \secref{sec:theory} and have thus been criticized~\cite{Lefever1995}.
A more realistic realization of externally maintained droplets obeys the thermodynamic principles and regulates the reaction mobilities.
In \figref{fig:ternary_active_droplets}, the external energy input~$\Delta\mu$ drives the conversion $B \rightarrow A$ inside the droplet, e.g., because an enzyme catalyzing this reaction segregates into the droplet~\cite{Kirschbaum2021}.
This particular model is know as the \emph{enrichment-inhibition model} since the enriched enzyme inhibits further droplet growth~\cite{Soeding2019}.

\subsubsection{Internally maintained droplets}
If the dilute phase is under-saturated ($\phi_B^{(0)} < \phiOut$), existing droplets dissolve and new droplets cannot emerge via nucleation.
Consequently, stable droplets can only form when an additional source of droplet material~$B$ is introduced.
The right column of \figref{fig:ternary_active_droplets} describes a specific example, where the chemical energy~$\Delta\mu$ drives the conversion $A \rightarrow B$ in localized regions (indicated by green areas in the plots).
If this local reaction is sufficiently strong, the fraction~$\phi^B$ can be pushed beyond the spinodal line, so  a droplet spontaneously forms in this region.
Since the surrounding dilute phase is under-saturated, the droplet continuously looses material to the dilute phase, where it is turned back into the precursor~$A$.
Larger droplets loose more material while the influx from the driven chemical reaction depends only weakly on droplet size.
Consequently, a stable droplet size emerges, which can also be controlled by the parameters of the reaction.
Note that the cyclic fluxes are opposite to the case in the previous section; see \figref{fig:ternary_active_droplets}D.
Since droplet material is created inside the droplet, we call this system \emph{internally maintained droplets}~\cite{Weber2019}.
It can be realized by localizing enzymes that catalyze the production of droplet material from precursor.
Since these enzymes induce droplet formation, this particular model is known as the \emph{localization-induction model}~\cite{Soeding2019}.
If the enzyme is not fixed in place but simply segregates into the droplet, this model leads to accelerate coarsening compared to traditional Ostwald ripening~\cite{Tena-solsona2019}.

\section{Future challenges}
\label{sec:future}

The study of phase separating systems that undergo active chemical reactions is in an early phase and many phenomena are yet to be discovered.
This section lists some aspects that are currently studied or will soon become important.
However, since the field is vast, this list is far from complete and I apologize if I missed relevant contributions.

\subsection{Multi-component fluids and complex reactions}

This review only discussed in detail binary fluids and a particular case of a ternary mixture.
However, experimentally relevant systems, and in particular biological examples, typically contain many more interacting components~\cite{Updike2009}. 
In fact, even ternary mixtures can display surprisingly complex phase diagrams~\cite{Hsu1974,Gonzalez-Leon2003,Bauermann2021,Devirie2021}  and this trend continues for increasing component counts~\cite{Mao2018,Mao2020}.
We can now simulate systems of a few tens of components~\cite{Shrinivas2021, Zhou2021a} and analyze large systems in particular cases using random matrix theory \cite{Sear2003,Jacobs2013,Jacobs2017,Shrinivas2021} and scaling analysis~\cite{deCastro2018}.
However, such random, unstructured interactions might not represent biological examples very well.
In fact, tuned interactions can lead to much more robust phase behavior~\cite{Zwicker2022,Jacobs2021}.
Components with specific interactions, e.g., in the form of surfactants, trapped species~\cite{Webster1998}, and solid particles adsorbing to interfaces~\cite{Ramsden1904,Pickering1907}, can also affect the coarsening dynamics that are inevitable in passive systems.
Taken together, this shows  we still lack a comprehensive theory of equilibrium states of multicomponent fluids.

Active fluids can exhibit even richer behavior than their passive counterpart.
Beyond the arrested coarsening that we showed in this review, specifically driven reactions can select certain phase behavior~\cite{Bauermann2021}, and result in hierarchical patterns~\cite{Tong2002a, Zhu2003}.
It will be interesting to see how other non-linear reaction schemes interact with the coarsening dynamics of phase separation.
A good starting point might be oscillating reaction-diffusion systems, e.g., based on the  Belousov-Zhabotinsky (BZ) reaction~\cite{Vanag2001} or the Brusselator model~\cite{Falasco2018}, which might show interesting spatio-temporal patterns once non-ideal diffusion related to phase separation becomes relevant.
In particular, the thermodynamic cost of converting chemical fuel into diffusive fluxes, quantified by  entropy production, needs to be investigated in more detail~\cite{Avanzini2021,Avanzini2020,Falasco2018}.

Studying phase separating systems with complex chemical reactions is a theoretical and experimental challenge.
The simplest theoretical approach couples phase separation to mass-action kinetics.
While such systems are typically easier to analyze, they violate thermodynamic constraints.
To connect to experiments, it is then crucial to also develop thermodynamically consistent theories.
Such a two-step approach unveiled the arrested coarsening, where models based on first-order reactions revealed stable patterns~\cite{Huberman1976, Puri1994, Glotzer1995, Christensen1996}, which were later substantiated by thermodynamically consistent stability analysis~\cite{Carati1997} and a full theory~\cite{Kirschbaum2021}.
A similar two-step approach might thus also be most promising for studying more complex behavior.
An alternative starting point would be thermodynamically consistent descriptions of reaction networks~\cite{Rao2016} to which phase separation could be added.

\subsection{Hydrodynamics and material properties}

The theory presented in this review neglects momentum fluxes, which would be described by a Navier-Stokes equation.
While the resulting effects are typically negligible in fluids with large viscosity, low viscosities might disrupt some behavior, like droplet division~\cite{Seyboldt2018}, and complex flow patterns could emerge in other examples~\cite{Lohse2020}.
For instance, surface tension gradients lead to Marangoni flow and complex dynamics of phase separation~\cite{Tan2016}.
While these systems are internally driven, an external drive can perturb the system even more strongly.
For instance, droplets break up in shear flow~\cite{Taylor1932,Rallison1984}, which selects a characteristic length scale even without reactions~\cite{Fielding2008,Stratford2007}.
It would be interesting to observe how chemical reactions alter such patterns.
Such complex dynamics including momentum fluxes might be best simulated using Lattice Boltzmann methods~\cite{Gsell2022,Krueger2016,Ledesma-Aguilar2014}, which would also allow studying complex geometries and turbulence.

Realistic fluids not only exhibit viscosity, but they often display complex material properties.
In particular, large macromolecules, e.g., in biomolecular condensates, can entangle, leading to long relaxation time scales~\cite{Jawerth2018,Jawerth2020}.
Moreover, droplets can be caught inside cross-linked networks or gels, which limits their coarsening dynamics~\cite{Fernandez-Rico2021,Lee2021,Ronceray2022,Style2018}.
More generally, every process that affects stresses locally impacts the pressure balance given by \Eqref{eqn:coexistence_many_P}, which underlies phase coexistence.
Consequently, strain stiffening environments limit droplet growth~\cite{Kothari2020, Wei2020}, and stiffness gradients bias droplets to softer regions~\cite{Rosowski2019,Rosowski2020,Vidal2020,Vidal2021}.
Since these material properties can also control the position and size of droplets, it will be interesting to study how chemical reactions augment the picture.

\subsection{Interaction with the environment}
The phase transition underlying phase separation provides a powerful process to sense external stimuli.
For instance, phase separation sensitively depends on global temperature~\cite{Choi2020}, which is for instance used to regulate sprouting in plants~\cite{Kim2021}.
Moreover, the dependence on ionic strength~\cite{Brangwynne2015}, salt, pH~\cite{AdameArana2020}, and  crowding agents~\cite{Andre2020} has been described in detail, and can be considered as multicomponent fluids where composition is a global control parameter.
Consequently, adjusting any of these parameters allows to control phase separation in the system.
One example is charging batteries where detrimental phase separation of lithium ions can be avoided by suitable protocols~\cite{Bazant2013, Bazant2017}.

External parameters can also be used to control the spatial details of a phase separating system.
One example are external potential gradients, like gravity or electric potential, which bias phase separation to one side of the system~\cite{Weber2017}.
In fact, the system's boundary plays a prominent role, since it might prefer one component in the fluid over another.
The resulting wetting and prewetting dynamics control phase separation~\cite{Zhao2021}, which is used in biology~\cite{Morin2022}.
Chemical reactions provide additional control via rate-controlling enzymes, which can be localized precisely.
Such spatial control over reactions can determine resulting patterns~\cite{Laghmach2021,Kuksenok2006} and localize droplets~\cite{Zwicker2014,Zwicker2018b}.
Finally, the interplay between bulk and surface dynamics implies a prominent role of geometry, which was described in detail for traditional reaction-diffusion systems, like the Min oscillations~\cite{Burkart2022,Brauns2021}, but could also control droplets, e.g., at the origin-of-life~\cite{Ianeselli2022}.
Spatially patterned catalysts could provide detailed control over the kinetics of chemical reactions, which in turn affect where droplets form.

\subsection{Nucleation and fluctuations}

This review only considered deterministic effects of phase separation and chemical reactions, but real system exhibit fluctuations from thermal noise and potentially other sources.
Since we derived phase separation kinetics using linear non-equilibrium thermodynamics, thermal fluctuations can be directly added to the dynamical equations using the fluctuation-dissipation theorem~\cite{Julicher2018,Cook1970}.
Capturing fluctuations in chemical reactions might be more challenging, although it might also not be necessary since diffusive fluxes dominate in most cases that we discussed here.
Including thermal fluctuations will allow to properly describe homogeneous and heterogeneous nucleation of droplets~\cite{Xu2014,Turnbull1950}.
Fluctuations will also be important in analyzing dynamics of single molecules, e.g., to compare to experimental measurements~\cite{Bo2021}.
While thermal fluctuations will often only weakly affect the dynamics, they may also have significant impact in particular parameter regions, e.g., via stochastic resonances~\cite{Gammaitoni1998}.

\section{Discussion}
This review demonstrated that chemical reactions can have a profound effect on phase separating fluids.
In equilibrium, reactions reduce the number of conserved quantities and thus the possible equilibrium states.
Consequently, binary systems only exhibit homogeneous states and ternary systems with a single reaction reduce to simple binary phase separation.
In systems with more components, chemical equilibrium selects a manifold in which equilibrium states are governed by phase separation of a reduced system.
Taken together, chemical reactions thus generally reduce the complexity of phase diagrams.

Chemical reactions become a powerful tool to control phase separation when they are actively driven, e.g., by providing fuel from the environment.
To affect phase separation, the energy of the chemical fuel needs to drive spatial fluxes.
This is possible when phase separation breaks the symmetry in the reaction networks, i.e., when the reaction fluxes depend on position or on local composition.
The spatial fluxes originating from these generic conditions can then suppress droplet coarsening, control droplet size, and also localize droplets.
Such systems fall in the large class of active matter since energy is consumed locally to affect dynamics.
However, since self-propulsion does not play a role and instead phases grow via diffusive fluxes, these systems are often classified as \emph{growing active matter}~\cite{Tjhung2020} in contrast to the more traditional motile active matter.

The combination of phase separation and driven chemical reactions provide unique properties that are useful in biology and technology.
The first-order phase transition underlying phase separation provides a sensitive response to changes in the environment and it allows to form distinct compartments without continuous energy input.
In contrast, the driven chemical reactions provide spatial-temporal control over diffusive fluxes and can thus affect where droplets form, how large they get, and how many there are.
The combination of both processes allows a complex regulation of spatial compartmentalization, which is sensitive to environmental conditions.
It is thus no surprise that such systems are ubiquitous in biology~\cite{Soeding2019,Yan2021,Henninger2020} and they will likely become more prominent in technological applications to control patterns in nano sciences, e.g., to produce structural color~\cite{Dufresne2009,Sicher2021} and other nanostructure.

\begin{acknowledgments}
I thank the entire \href{www.zwickergroup.org}{Zwicker group} for stimulating discussions and I particularly thank Francesco Avanzini, Massimiliano Esposito, and Jan Kirschbaum for a critical reading of the manuscript and helpful comments.
Funding by the Max Planck Society is gratefully acknowledged.
\end{acknowledgments}

\appendix

\bibliographystyle{elsarticle-num-names}
\bibliography{references}%

\end{document}